\documentclass[12pt,preprint]{aastex}

\begin{document}
\title{A Universal Central Engine Hypothesis for Short and Long GRBs}
\author{ David Eichler \altaffilmark{1}, Dafne Guetta
\altaffilmark{2}, \&
 Hadar Manis\altaffilmark{1}
 }
\altaffiltext{1}{Physics Department, Ben-Gurion University,
Beer-Sheva 84105, Israel; eichler@bgu.ac.il}
\altaffiltext{2}{Osservatorio astronomico di Roma, v. Frascati 33,
00040 Monte Porzio Catone, Italy.}
\begin{abstract}
It is noted that X-ray tails  (XRTs) of short, hard $\gamma$-ray
bursts (SHBs) are similar to X-ray flashes (XRFs). We suggest a
universal central engine hypothesis, as a way of accounting for this
curiosity, in which SHBs differ from long $\gamma$-ray bursts (GRBs)
in prompt emission because of the differences in the host star and
attendant differences in the environment  they present to the
compact central engine (as opposed to differences in the central
engine itself).
 Observational constraints and implications  are discussed,
especially for confirming putative detections of gravitational waves
from merging compact objects.
\end{abstract}

key words: $\gamma$-rays: bursts

Short $\gamma$- ray bursts (GRBs), originally defined to be GRBs
lasting less than 2 seconds and predicted to be a separate class of
phenomena (Kouveliotou et al 1993), are now widely suspected of
being two merging compact objects. This somehow distinguishes them
in their duration and other properties from the core collapse of a
massive star. In the former case, primary emission can in principle
be detected even if it comes from less than 2 lightseconds from the
black hole, which would probably be impossible for long bursts,
where envelopes would obscure photon emission from these scales. The
central engines of each, however, are likely to be similar: a black
hole of maximum angular momentum surrounded by an accretion disk of
matter near nuclear densities.

A fundamental open question about short GRBs is why they are so
short. Is it because the central engine (presumably a black hole fed
by an accretion disk) operates on a shorter timescale than those of
long GRBs? Alternatively, it may be the timescale over which the
$\gamma$- rays are visible. For example, the prompt $\gamma$- rays
could be scattered off slow baryons (Eichler and Manis 2007) and the
short duration is a result of the baryons getting accelerated to a
high enough Lorentz factor to exclude much further emission along
our line of sight.    What seems like  a short burst to us would
then appear to be a much longer burst to some other observer along a
different line of sight. The hypothesis can account for the fact
that short bursts tend to have harder spectra, lower luminosity, and
a larger solid angle than long bursts. It also accounts for the
inverse correlation between luminosity and spectral lag (Hakkila et
al. 2008 and references therein) in a simple manner, assuming the
acceleration is due to the primary radiation pressure.

In this letter we propose a universal central engine hypothesis for
both short and long bursts.  We suggest that the same compact,
central engine can produce what seems to be a  short, hard burst to
a viewer at  large viewing angle, and a long burst to a viewer at
smaller viewing angle. The distinction  between "central engine" -
by which we mean the post-collapse compact object - and "progenitor"
is emphasized. We keep the now common view that short hard bursts
come from merging neutron stars or other systems without a large
envelope, whereas the long bursts typically come from progenitors
with large, post main-sequence envelopes.\footnote{The differences
between the progenitors of short and long GRB, might, {\it a
priori}, be suspected of giving rise to different types of central
engines; our point is simply that such differences do not appear to
be crucial or necessary for understanding the systematic differences
between short and long bursts. Similarly, the differences in the
host environment of the central engines might give rise to
differences in the collimation of short and long bursts, and this
 might indeed have some observational consequences, but neither is this the
 main point of this letter.  Here we merely note that the observed differences
 can be accounted for by  different viewing angle.}

 That typical viewing
angle should correlate with the progenitor and host environment is
straightforward: a central engine sitting within a massive envelope
is obscured by the envelope, and fireball material within the
envelope can be observed directly down the hole bored by the
fireball in the envelope (Figure 1). Observers outside the opening
angle $\theta_o$ established by this hole can see emission from
material only after it has nearly emerged from the envelope, and,
from well outside the opening angle of the emerging material
(including its $1/\Gamma$ emission cone), they detect kinematically
softened emission that is nearly backwards in the frame of the
fireball. Such an  observer cannot view matter while it is well
within the confines of the envelope ($R\le 10^{12}$cm).
 When, on the other hand, there is no
massive envelope obscuring the central engine, matter can be seen
from  within $10^{12}$cm, and subsecond timescales become possible.
The significance of the host/progenitor may then be the angles at
which it allows the burst to be observed, and, therefore, the stage
of the burst that is observable.

The unified model proposed here should be contrasted to that of
Yamazaki et al (2004), which, somewhat presciently, was made {\it
before} the operation of Swift (and the resulting localization of
SHBs). They proposed that  short bursts come from the same central
engine as long bursts, and that they appear short when one emitting
"minijet" of many comes close enough to the line of sight as to
dominate over the contribution of all the others. Observers at large
offset angle to the axis of the swarm of minijets see X-ray flashes,
but little or no hard emission. In the Yamazaki et al picture, it
would be hard to understand why SHBs typically come from different
types of galaxies and, by inference, different types of progenitors.
 It would also be hard to understand why
SHBs are {\it underluminous} (or overly hard) in the context of the
Amati relation. It is not clear that the hard part of the GRB would
always precede the soft part.  Finally, it would be hard to
understand why the small scale time structure and spectral lags in
SHBs are qualitatively different from those in long GRBs. In the
unified model we propose here, on the other hand, the hard photons
of the SHB are seen at large viewing angles, but from an earlier
stage of the fireball's acceleration, and these observations follow
naturally.

Below, we summarize the observations that motivate the universal
central engine hypothesis. We then show that a particular model for
GRB subpulses can produce a viable model for SHBs and the X-ray
tails.

{\it Observational Motivations:} SHBs frequently  display long X-ray
tails that compare in duration to long X-ray flashes. The discovery
(e.g. Donaghy et al 2006; Norris and Bonnel, 2006; Gehrels et al
2006) confirmed by Swift  that short bursts have X-ray tails (XRTs)
of much longer duration than the burst itself heightens the
suspicion that the central engine continues to operate for longer
than 2 seconds. Donaghy et al report that most SHBs observed with
HETE II (which has a lower photon energy threshold than Swift) have
long, soft tails, whereas the fraction of Swift SHBs is somewhat
less, about half. We may interpret this as XRTs being slightly
softer than XRFs and/or as Swift being more sensitive than HETE II
to the short, hard phase of SHBs. We argue below that this is
expected, because the larger the viewing angle, the shorter, harder
the emission of the short phase, when by hypothesis $\Gamma \sim
1/\theta$, and the softer the tail emission when $\Gamma$ has
reached its terminal value.

  Apart from possibly being slightly softer,  these XRTs are quite similar
to $\gamma$-ray-silent  X-ray flashes (XRFs), which have been
proposed to be "off axis" GRBs. This interpretation of XRFs has also
been supported by their tendency to show depressed X-ray afterglow
(relative to normal GRBs) until $\sim 3 \times 10^5$s after the
prompt emission (Sakamoto et al 2008), after which the afterglow
appears to be of about the same intensity as that of a classical
GRB. This suggests that we are seeing a kinematically suppressed,
under-blue shifted signal that is predicted for an offset viewer.

There exists by now some evidence that SHBs are beamed into a small
solid angle, similar to long GRBs. Fox et al. (2005) interpreted the
steepening of the optical afterglow light curve of GRB 050709 and
GRB 050724 in terms of a jet break, translating into a beaming
factor $f_b^{-1}\sim 50$ (with $f_b$ the fraction of the $4\pi$
solid angle within which the GRB is emitted). Soderberg et al (2006)
found a beaming factor of $\sim 130$ for GRB 051221A. Therefore,
with the present data, the beaming angle of SHBs seems to be
 in a range of  $\sim 0.1-0.2$ radians. The  question is
 how this beam width compares to that of the X-ray tail.

Below,  we  summarize the data on SHBs, note that the XRTs (unlike
the hard emission) obey the Amati relation (Amati et al. 2002),
describe a particular version of a universal central engine
hypothesis,  and attempt a rough estimate of the beam width of the
XRTs  of SHBs based on the supposition that they are basically XRFs.

 We have considered all the short bursts reported
by Swift from its launch (November 2004) until March 2008;  this
constitutes a sample of 28 bursts. In Table 1 we report the observed
data relative to those for which X-ray emission was detected by the
X-ray telescope on board  Swift. As noted by Donaghy et al (2006),
prolonged soft emission in HETE II data is a rather reliable
signature of SHBs, but it  is clear that there is large scatter in
the luminosity of the XRTs  relative to the $\gamma$-ray luminosity,
and, in the Swift data, XRTs are {\it not} a reliable indicator for
a SHB.

The XRTs  are quite similar to $\gamma$-ray-silent XRFs detected by
HETE-II, BeppoSax and Swift (see Figure 2). We made a spectral
analysis of all the XRTs that could be detected by the WFC and did
not find any evidence of a spectral break in the band 0.3-10 keV. We
have therefore assumed a spectral peak at $\sim 10$ keV with an
uncertainty of $\sim 8 $ keV. The only burst that seems to have a
higher energy break is 050724 (Campana et al. 2006) and we took
$E_{\rm peak} \sim 20 $ keV  for this burst. We see that the XRTs
obey the Amati correlation to within the uncertainties, whereas the
prompt $\gamma$- ray emission is far removed from this correlation.

{\it Interpretation:} That XRTs of SHBs are consistent with the
Amati relation is what is expected if both the short hard emission
and the X-ray tail are attributed to the offset viewing of what
might be observable as a classic long GRB from a different
direction. The X-rays are photons beamed backward in the frame of
the classical fireball. They are reduced in frequency as the first
power of the Doppler factor, and, in time-integrated fluence, by the
square of the Doppler factor, when the beam is wider than the
angular separation $\theta_V$ of the observer's line of sight from
the beam (Eichler and Levinson 2004). This gives the Amati
correlation. (When, on the other hand, $\theta_V$ is
 comparable to or larger than $\theta_o$, the
 apparent luminosity decreases as a steeper power of peak frequency [Yamazaki et al 2002, 2004,
 Eichler and Levinson 2004]).
 The hard photons, on the other hand, are scattered into the line of
sight  by baryonic material that has not yet been accelerated beyond
a Lorentz factor of $1/\theta_V$ (but soon will be). As such, the
primary luminosity, as seen by observers in the beam, is diluted by
the scattering, because the $1/\Gamma$ cone at low $\Gamma$ is much
wider than it will end up when maximum Lorentz factor is reached.
Whereas the scattering reduces only modestly the individual photon
energies to observers within $1/\Gamma$ of the axis of scatterer's
motion [$cos\theta_V \ge \beta$] - $E^{\prime}/E$ being at least
$1/\Gamma^2(1-\beta^2)(1+\beta)\ge 1/2$ - it greatly dilutes the
fluence by spreading it over a much wider solid angle. So the
fluence is greatly below what the Amati relation would predict for
viewers that are within $1/\Gamma_a$ of the primary beam when it is
finally at its terminal Lorentz factor $\Gamma_a$. The observer at
wide angle (from the direction of the low $\Gamma$ scatterer) sees a
much shorter hard pulse  than the viewer at smaller offset angle
(Eichler and Manis 2007),  as shown in Figure 3, because the
acceleration time as measured by the observer is proportional to
$\Gamma$. In these figures we plot the light curve of the scattered
emission from a single accelerating cloud with point-like geometry
as seen by viewers at two different viewing angles. (We stress that
this is not the same as predicting a light curve from the burst,
which has a finite solid angle and time interval in which scatterers
can be injected.  The data superimposed in the same graph is merely
for reference.) Other factors could contribute to the duration, such
as the intensity of radiation pressure that causes the acceleration.

 Assuming a Lorentz factor for the blast of $\Gamma(t)
\simeq 100 (t/100s)^{-3/8}$ (Sari et al. 1998),
 we estimate the Lorentz factor of
the blast wave after $3 \times 10^5$s, the typical recovery time for
XRFs (Sakamoto et al 2008), to be $\Gamma(10^{5.5}s )\sim
(10^{11/16}) \sim 5$.  Attributing the afterglow recovery to the
decrease of the blast's Lorentz factor down to $1/\theta_V$, one
estimates that XRFs are typically observed at an offset angle of
$\theta_V \sim 0.2 \sim 10^o$ from the blast. In fact, a complete
recovery requires that $\Gamma$ decline to comfortably below
$1/\theta_V$, so $\theta_V$, defined to be the angular distance to
the edge of the jet, is better estimated to be less than 0.2.
 Writing $\theta_V$ as $10^{-1}\theta_{-1}$, the typical spectral
peak of the XRTs  $E_{peak}$ as $30 E_{x,30}$KeV, and the expected
spectral peak
 measured by the head-on observer as
  $ E_{ho} \times 1 $Mev, we
  estimate the Lorentz factor of the fireball that emits
the prompt emission to be given by $0.03E_{x,30}/E_{ho}=1/\Gamma^2
(1+\beta)(1-\beta cos\theta) \sim (\theta_V\Gamma)^{-2}$, or
\begin{equation}
\theta_V\Gamma \sim 6 [E_{x,30}/E_{ho}]^{1/2}.
\end{equation}
For $\Gamma \sim 10^2$, and $E_{x,30}, E_{ho}$ both $ \sim 1$, this
gives $\theta_V \sim 6 \times 10^{-2}[E_{x,30}/E_{ho}]^{1/2}$, which
is consistent with the estimate of $\theta_V$ from the afterglow
recovery time. Assuming that the jet itself has an opening angle of
order $10^{-1}$ radians, this gives an opening angle for an XRF of
about 0.1 to 0.2 radians,  in reasonable  agreement with the
estimate from the flat phase of the afterglow. That the offset
$\theta_V$ is comparable to the opening angle of the fireball jet
$\theta_o$ suggests that for $E_{30}\ll 1$, the luminosity of the
XRF should drop below that predicted by the Amati relation.

That extended soft emission is a reliable indicator for  SHBs
(Donaghy  al. 2006) suggests that the solid angle in which the soft
photons are detectable by HETE II  is at least as large as that from
which the hard $\gamma$-beam is detectable. On the other hand, the
large variation in X-ray to $\gamma$-ray fluence suggests that we
should be cautious about making simple generalizations regarding the
relative characteristics of the X-ray tail and the $\gamma$-ray
beams. Given our estimates of 0.1 to 0.2 radians for both the soft
and hard beams, we could attribute the large variation in hard/soft
emission ratio to the fact that the opening angles of the XRTs  and
hard $\gamma$-ray emissions are comparable, and that one can be
observed near the ragged edge of the other. In our model, moreover,
the fraction of hard $\gamma$- rays scattered into our line of sight
can be highly variable from one burst to the next. Furthermore,  the
directions of the prompt $\gamma$-radiation and the accelerating
baryons need not be the same (e.g. Eichler and Granot 2006, and
references therein), so the respective relations of the observer's
line of sight to each of them is a somewhat free parameter. This
affects the Lorentz factor of the scatterer that contributes to our
line of sight, and hence the extent of solid angle dilution of the
photon intensity. XRTs should therefore be considered as an
important  complement to (but not necessarily better than) SHBs in
corroborating LIGO signals.

Long GRBs, XRFs, SHBs and XRTs each have two parameters - their
cosmic event rates per unit volume and their beaming factors - for a
total of eight parameters. Measuring the relative detection rates
and distribution of distances of each of the four categories of
events reduces this to four free parameters. The universal central
engine hypothesis, in its simplest and most naive form, together
with the offset viewing hypothesis for XRFs  posit that a) the rate
per unit volume of XRTs  is the same as that of SHBs, b) the
physical parameters of XRFs and XRTs are the same, c) the relative
event
 cosmic rates of XRFs and XRTs per unit volume are the same as for long vs. SHBs, and
d) the rate per unit volume of XRFs is the same as for classical
long GRBs. These are only four assertions that constrain the four
unknowns. So, although it  may be possible to constrain the
parameters of such a beaming factor within the framework of a
universal central engine hypothesis via observations, the above
considerations do not overconstrain the model enough to test its
validity.

On the other hand, additional information could provide further
tests. There may be small differences between XRTs and XRFs imposed
by the different types of host stars, differences in their
subsequent afterglow patterns as well as information on the host
galaxies. Further into the future, a viable data set of LIGO events
would allow a test of the relative beaming factors of SHBs and XRTs.
We suggest that LIGO should operate together with efficient SHB
detectors and wide field X-ray cameras. This would not only improve
the chances for corroborating LIGO detections of mergers, but would
enable these detections to teach us more about the associated high
energy processes as well.

Our suggestion that some XRTs  of SHBs are XRFs, {\it combined with}
 the hypothesis that they correspond to offset viewing of
 a long burst in some
other direction, predicts that a large enough sample of XRFs, even
if unbiased by any $\gamma$-ray trigger, should have a subset that
correlates with SHBs. A careful analysis, however, shows that BATSE
should have detected less than one SHB coincident with any X-ray
flash. A larger sample of XRFs detected while a SHB detector is
operating would give tighter constraints.

{\it Further Consequences:} Short bursts, inasmuch as they are
believed to be merging compact objects  (neutron star-neutron star,
NS-NS, or neutron star-black hole, NS-BH), are expected to be
closely connected to gravitational wave signals, and potential
candidates, if close enough, for detection by LIGO.  The horizon of
first generation LIGO and Virgo for NS-NS, NS-BH mergers is $\sim
20$ and $43$ Mpc, respectively, while advanced LIGO/Virgo  should
detect them out to a distance of $\sim 300$ and  $650$ Mpc (for a
review see Cutler \& Thorne 2002).
  Guetta and Stella (2008) have recently estimated that, assuming a beaming
  factor $\sim 100$,
 a sizeable fraction of  gravitational wave events detectable by LIGO II is expected
 to be coincident with SHBs, which provides a new,
interesting perspective for the Advanced LIGO/Virgo era. Here we
note that a wide angle X-ray camera in addition to a Swift-type
detector that triggers on hard $\gamma$-ray emission could possibly
increase our ability to corroborate LIGO signals as well as learn
more about merger events. As these events could be of marginal
statistical significance, it would be good to verify them
independently with detections of high energy emission that is
believed to be associated with mergers. The complex,  multicomponent
nature of SHBs suggests that careful thought should be given as to
the best way to corroborate putative gravitational wave events.

{\it Summary:} We  have suggested that short, hard GRBs may have the
same central engine as that of long GRBs, though having a different
size host envelope, and that their duration is determined by the
acceleration time of a relativistic scatterer that scatters them
into our line of sight. A very rough estimate of the opening angle
of XRTs, based on their hypothesized similarity to XRFs, is
 0.1 to 0.2 radians, which is comparable to estimates of the
opening angles for the hard emission. It is thus difficult to say
which would be better for corroborating nearby compact mergers
following LIGO triggers. Further information on the relative
detectabilities of XRTs  and the corresponding short hard
$\gamma$-emission could be obtained by a wide field X-ray camera and
$\gamma$-ray detectors working together. In six of 12 cases with
known redshifts, the X-ray tail would exceed $10^{-10}$erg/cm$^2$s
had the burst been within 300 Mpc (distance of GW detectability if
the SHBs come from NS-NS mergers), so that LIGO might in such cases
act as the primary trigger, which would have the valuable property
of being free of any electromagnetic spectral bias.

One interesting prediction of the universal central engine
hypothesis is that, while much of the X-ray fluence comes after the
location and slewing of the Swift X-ray telescope,  much of the
fluence also can come out on  a short timescale, the so called
"spike", depending on whether the source has a soft primary
component. This appears to be consistent with HETE-II observations
of short bursts (Donaghy et al 2006, figures 2-5,13, table 7). An
important distinction should be made between
 the short rise of soft emission and the longer tail.
 The "spike" (i.e. rise and peak)  consists either of a) photons that were soft at the primary
 source, and lost only about half their
 (observer frame) energy during the scattering
 or b) primary radiation
 that was softened by reprocessing at the scatterer, while the
 tail consists of photons that may  well have been
 hard at the source and were drastically softened in the observer's
 frame by making a near 180 degree
 rear end collision with the scatterer. The former should have a
 rise time that is nearly energy independent, as it is established by the
 acceleration time of the scatterer to $1/\theta $, while the latter
have a much longer decay time at low energy because the
collision-softened photons populate the low energy bins at large t
 (Eichler and Manis 2007).
Thus, the relative shapes of the light curves in different energy
bins, which reflect the relative contributions of these two classes
of photons to the X-ray tail, can be very sensitive to the primary
spectrum, which, in turn,  can be sensitive to the type of
progenitor/host. For example, if the prompt, soft photons have a
component of thermal emission from the back side of the scatterer
that has been heated by Compton recoil, we expect this component to
have the same time profile as the hard $\gamma$-rays. Such a
component could be more important for SHB environments, where the
scatterer is closer to the central engine and sees a stronger
radiation field. Moreover, at lower Lorentz factor, its back end
sees a harder radiation field, so the Compton recoil heats more
efficiently. This and several other matters beyond the scope of this
paper bear further investigation.

A serious quantitative model of a SHB light curve within the context
of the ideas sketched here should take into account the following:
a) The scatterer may in fact be a swarm of individual clumps running
into each other and accelerating uniformly only as a group. Figure 3
should therefore be taken, at best,  as a rough envelope that
characterizes the general trend of the light curve. The individual
subpulses may be scattered photons from scatterers that are
accelerating (as well as decelerating) on a somewhat faster
timescale,  and therefore have smaller positive (as well as
negative) spectral lags.  b) The distribution of scatterers as a
function of angular separation from the observer's line of sight is
unknown but there are probably more at larger angles. The observed
light curve is the sum over the individual contributions from the
members of the swarm. c) While the scatterer is at modest Lorentz
factor, backscattered photons may be intercepted by pair-producing
collisions with primary photons. This can lower the X-ray luminosity
near and just after the peak, when, for viewers at large angle, the
Lorentz factor is still modest. (Eventually, as the Lorentz factor
picks up, the primary photons in the frame of the scatterer are not
energetic enough for pair production.) Thus, while primary, soft
photons may escape immediately, the nearly 180 degree backscattered
hard photons escape only after the overall Lorentz factor of the
scatterers is high enough.

The universal central engine hypothesis also predicts that,
occasionally, we are close to the axis of a GRB that originates from
a SHB-type host. In such a case, the GRB would appear long in
duration, but of the "short" variety in other ways. Indeed, there
are such bursts (e.g. 060614) that confound a simple two class
classification scheme (Donaghy et al 2006; Gehrels et al 2006).

\acknowledgements
We thank  Luigi Piro and Matteo Perri for useful discussions. We acknowledge support
from the U.S.-Israel Binational Science Foundation, the Israel
Academy of Science, and the Robert and Joan Arnow Chair of
Theoretical Astrophysics.

\clearpage

\begin{table}[t]
\caption{Properties  of SHB  prompt and afterglow emission as
detected by Swift  X-ray telescope and HETE-2 (indicated with the
*);
 the + indicates that
they could be detected by the WFC. The X-ray flux is estimated at
60-100 sec after the burst and is given in the 0.3-10 keV energy
range. In the last column we report what the X-ray flux would  be if
the SHB were at a distance of LIGO (advanced version)  detectability
(300 Mpc if SHBs come from NS-NS mergers). }
\begin{tabular}{lllllll}
 \hline
GRB & $z$ & $S_{\gamma}$ & $E_{\gamma, iso}$ & $F_x   $& $ E_{x,iso}$
& $F_x   $ (@ 300 Mpc) \\
   &    & $10^{-7}$ erg/cm$^2$  & $10^{49}$ erg &  $10^{-11}$ erg/cm$^{2}/s$& $10^{49}$ erg
   & $10^{-11}$ erg/cm$^{2}/s$\\
  \hline
 050709*$^+$&0.16  & 3$\pm$ 0.38   &  1.4     & 800 & 3.4  & 3.3 $\times 10^3$ \\
 050724$^+$ & $0.258$    & 6.3 $\pm$ 1    & 7.2        & 1200 & 10.1 &9.9 $\times 10^3$\\
 051210$^+ $&                     & 1.9$\pm$ 0.3 &           & 90 &  & \\
 051221$^+$ & 0.546        & 22.2 $\pm$ 0.8 & 84        & 20  & 0.6 & 590\\
 060313 $^+$&                    & 32.1 $\pm$ 1.4 &         & 30&  &\\
  071227$^+$  & 0.383 &    2.2 $\pm$ 0.3   & 4.0 & 46& 0.87 &  854\\
  050509B & 0.225& 0.23$\pm$ 0.09& 0.2 & 0.06 & 4.5$\times 10^{-4}$  &0. 44 \\
 050813 & 0.7& 1.24 $\pm$ 0.46 & 5.2 & 0.6& 0.025 & 25\\
 050906 &      & 0.84 $\pm$ 0.46&   & $<0.007$ &     &\\
 050925 &     & 0.92 $\pm$ 0.18 &    & $ <0.003$  &    & \\
 060502B& 0.287 & 1 $\pm$ 0.13 & 1.15 & 0.1 & 0.001 & 0.98\\
 060801 &       & 0.8 $\pm 0.1$&    & 0.1 &   &\\
 061201 & 0.11 & 3.3 $\pm$ 0.3 & 0.7 & 10 & 0.02& 24\\
 061217 & 0.827 & 0.46 $\pm$ 0.08 & 2.4 & 0.1 & 0.005 &4.9\\
 070429B & 0.904 & 0.63 $\pm 0.1$ &3.5 & 0.11& 0.006 & 5.9\\
 070724 & 0.45 & 0.3 $\pm 0.2$ & 0.6 & 0.05 & 0.0012 &1.2\\
 070729 & 0.904 & 1.0 $\pm$ 0.2&  5.6   & 0.024 &   0.001& 0.98\\
 070809 &             & 1.0                  &        &  0.179 &         & \\
 071112B &                        &  0.48                   &      &$<0.02$&  &\\

 \hline
\label{t:fit}
\end{tabular}
\end{table}

\clearpage

\begin{figure}
\includegraphics[width=1.0\columnwidth, keepaspectratio]{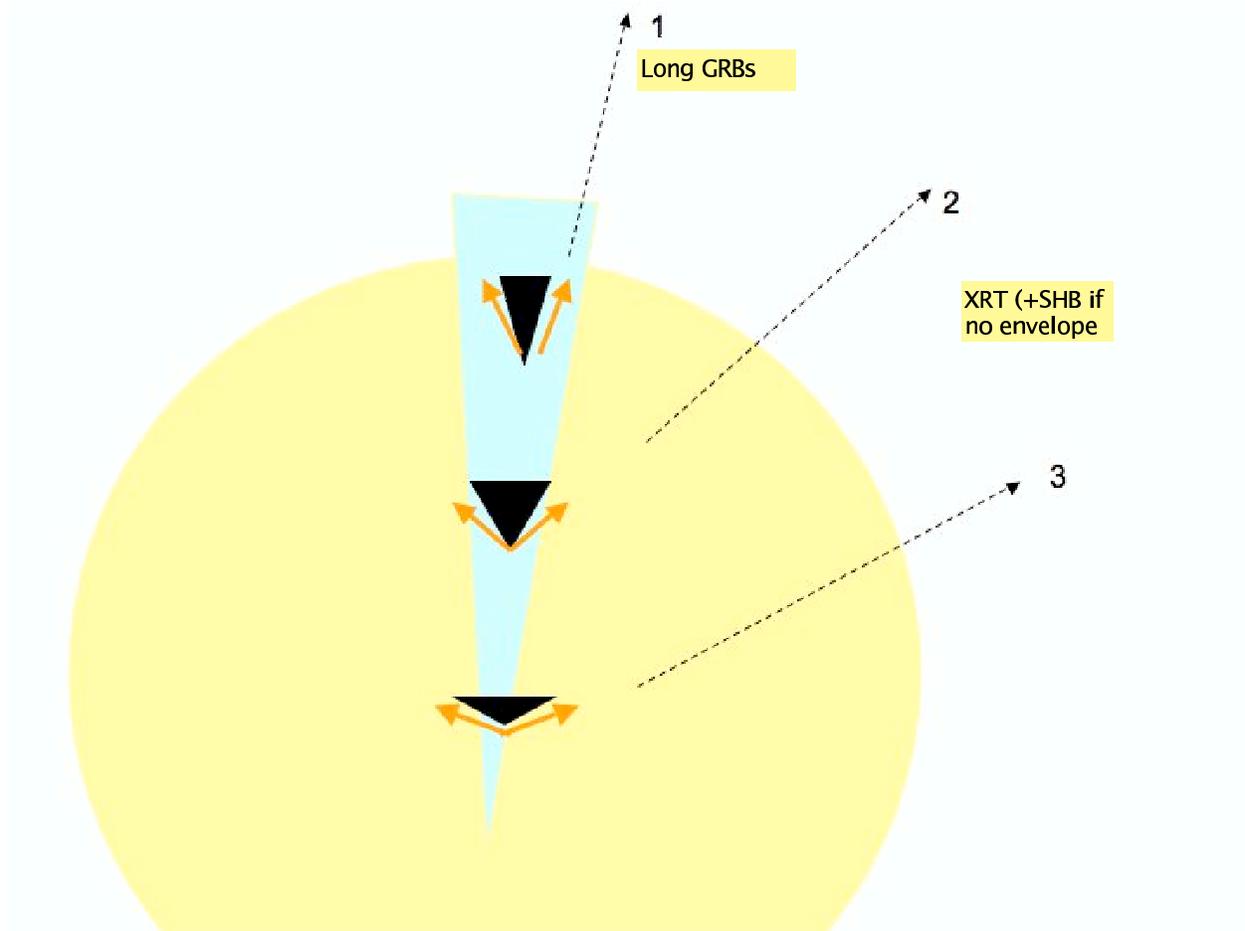}
\figcaption[FileName]{\label{come}   A schematic drawing of the
output of a "universal central engine". The black cones depict the
shadow cast by an optically thick scatterer that scatters or emits
radiation from its back end. The three positions of the scatterer
represent its increasing Lorentz factor $\Gamma$ as it is
accelerated by radiation pressure. Observer 1 sees a long
$\gamma$-ray burst, emitted at large Lorentz factor. Observer 2 sees
an XRT, and, only if there is no giant envelope obscuring the line
of sight, can also observe a short, hard burst of scattered
$\gamma$-radiation as  the scatterer accelerates through a Lorentz
factor that is the reciprocal of the observer's viewing angle.
Observer 3 sees an even weaker, softer XRT, and (also only if his
line of sight is not obscured) an even shorter hard GRB. }
\end{figure}

\begin{figure}
\includegraphics[width=1.0\columnwidth, keepaspectratio]{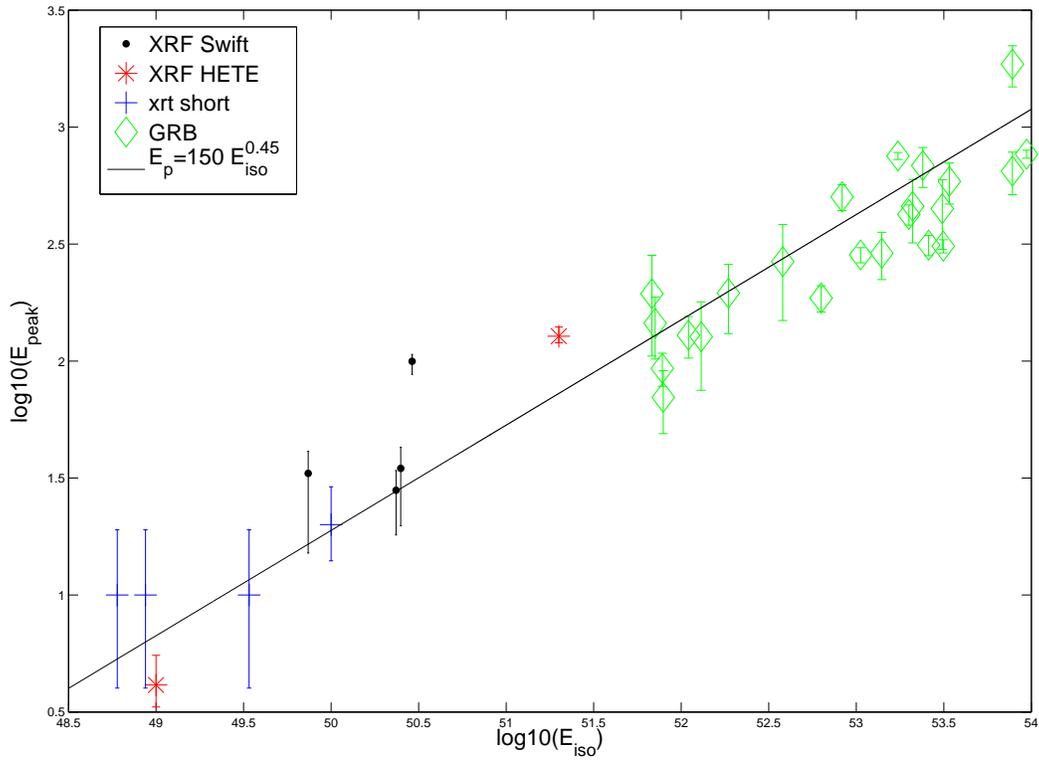}
\figcaption[FileName]{\label{cdf}  $E_{\rm peak}$ and $E_{\rm iso}$
values for the XRF detected by HETE and Swift and for the X-ray tail
of SHBs.  We also plot the relation for normal GRBs (Amati et al.
2002).}
\end{figure}

\begin{figure}
\includegraphics[width=0.7\columnwidth, keepaspectratio]{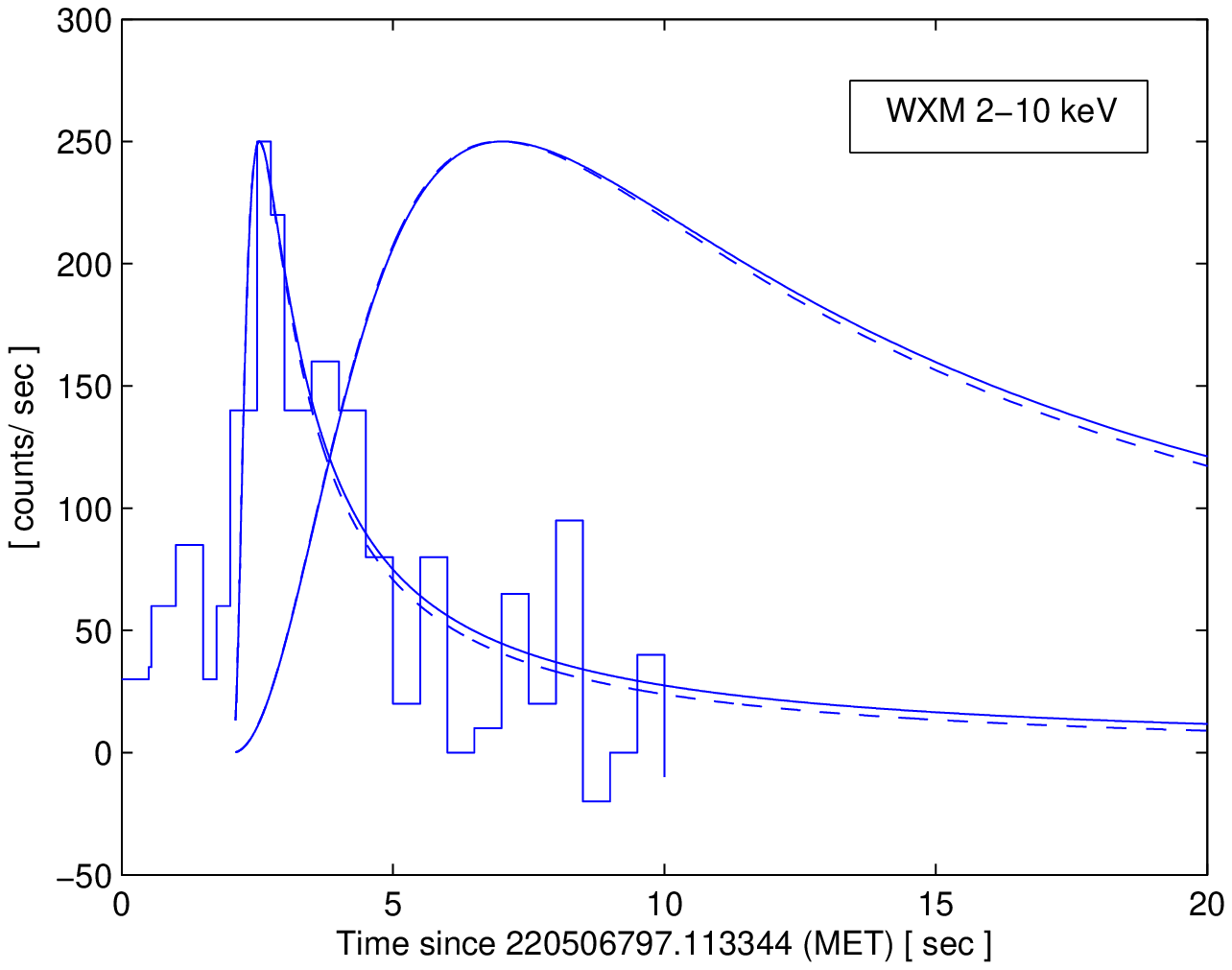}
\includegraphics[width=0.7\columnwidth, keepaspectratio]{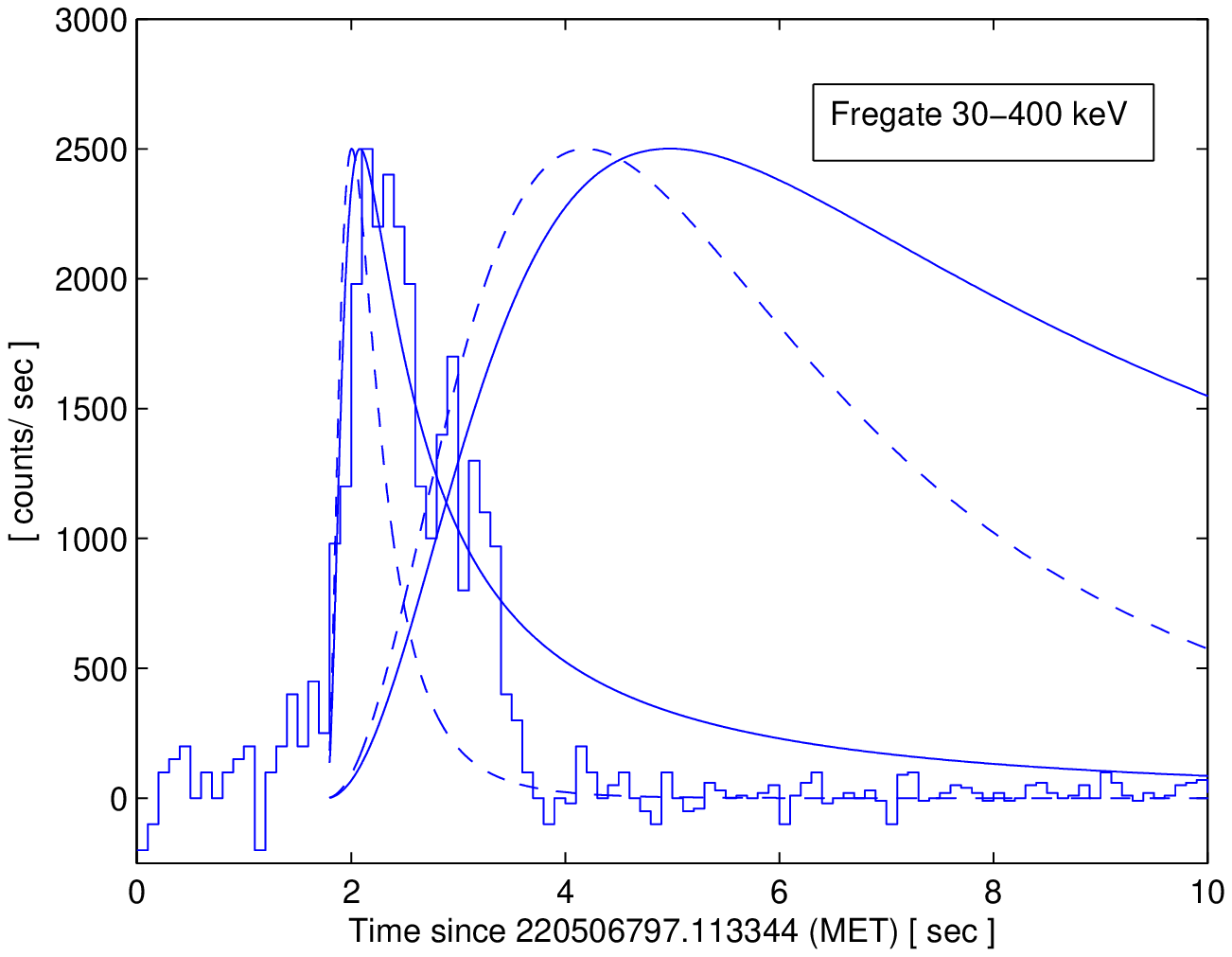}
\figcaption[FileName]{\label{f01}\footnotesize Upper panel: The
light curves at 2 (solid line) and 10 KeV (dashed) are shown for
observers that detect scattered emission off a single point-like
scatterer that undergoes constant acceleration, at viewing angles
$\theta_V$ of 0.02   and 0.2 radians  from the direction of the
acceleration. Lower panel: The same but for the photon energies 30
(solid) and 400 KeV (dashed). In each case the broader light curve
is for the smaller viewing angle 0.02, and the apparent rise time is
 inversely proportional to sin$\theta_V$.  The primary spectrum is
taken to be $E^{-1/2}e^{-E/600\,KeV}$. The plotted data is  GRB
060121 as observed by the HETE II WXM (2-10 KeV) and Fregate (30-400
KeV) where the x axis is labeled in seconds. The theoretical curve,
if it has a rise time of $t_0$ seconds, would  correspond to an
acceleration  (expressed here as an equivalent radiative force) in
the instantaneous rest frame of the scatterer
 of $d\beta^{\prime}c/dt^{\prime}$=
$\frac{1-\beta}{1+\beta}(\sigma_T/\tau m_pc)\frac{L}{4\pi r^2}\sim
t_0^{-1}(sin\theta)^{-1}(\sigma_T/\tau m_pc)\frac{2.5\times
10^{46}}{4\pi \times 10^{24}}\frac{erg}{cm^2s}$, where $\sigma_T$ is
the Thomson scattering cross section, and $m_p$ is the mass of the
proton (EM07). For SHBs, the theoretical curve might be more
appropriately scaled to subpulses. The normalization of the
theoretical curves is arbitrary for convenience of plotting. The
absolute normalization will be discussed elsewhere.}
\end{figure}

\end{document}